\documentstyle[aps,prl,psfig]{revtex}
\begin{document}

\title{Specific heat and thermal conductivity in the mixed state of MgB$_2$}
\author{L. Tewordt and D. Fay}
\address{I. Institut f\"ur Theoretische Physik,
Universit\"at Hamburg, Jungiusstr. 9, 20355 Hamburg, 
Germany}
\date{\today}
\maketitle
\begin{abstract}
       The specific heat $C$ and the electronic and phononic thermal
conductivities $\kappa_e$ and $\kappa_{ph}$ are calculated in the
mixed state for magnetic fields $H$ near $H_{c2}\,$.  The effects of
supercurrent flow and Andreev scattering of the Abrikosov vortex lattice
on the quasiparticles are taken into account. The resulting function
$C(H)$ is nearly linear while $\kappa_e(H)$ exhibits an upward
curvature near $H_{c2}\,$. The slopes decrease with impurity scattering
which improves the agreement with the data on MgB$_2\,$.  The ratio
of phonon relaxation times $\tau_n/\tau_s = g(\omega_0,H)$ for
phonon energy $\omega_0$, which is nearly a step function at
$\omega_0 = 2\Delta$ for the BCS state, is smeared out and tends to
one for increasing $H$. This leads to a rapid reduction of 
$\kappa_{ph}(H)$  in MgB$_2$ for relatively small fields due to the
rapid suppression of the smaller energy gap.
\end{abstract}
\vspace{0.5in}
    Numerous experiments indicate that the superconducting state of 
MgB$_2$ ($T_c = 40$K) \cite{Nagamatsu} is a conventional s-wave pairing 
state mediated by the electron-phonon interaction. Nevertheless, the
theoretical explanation of physical quantities is demanding because two 
different energy gaps are formed where the larger gap $\Delta_1$ is
associated with the nearly cylindrical $\sigma$-sheets of the Fermi surface
and the smaller gap $\Delta_2$ with the three-dimensional $\pi$-sheets. 
Analysis of the temperature dependence of the measured thermal 
conductivity in the basal plane of  MgB$_2\,$ \cite{Sologub1} gives 
supporting evidence for two different gaps on different Fermi surface
sheets. The situation is complicated because the thermal conductivity 
$\kappa$ is a sum of an electronic part $\kappa_e$ and the phononic or
lattice conductivity $\kappa_{ph}$. In Ref.~\onlinecite{Sologub1} $\kappa_e$ 
and $\kappa_{ph}$ were calculated with help of the BCS theory \cite{BRT} 
where  $\kappa_e$ is limited by elastic scattering (scattering by phonons is
omitted) and $\kappa_{ph}$ is limited by electron scattering. This theory for
$\kappa_{ph}$ is supplemented by adding to the phonon relaxation rate due
to scattering by electrons, the relaxation rates due to scattering by point
defects \cite{KlemTew} and other defects. We remark that the peak in 
$\kappa_{ph}$ below $T_c$ in the cuprate superconductors has been fitted
by including scattering by sample boundaries, sheet-like faults, and 
dislocations \cite{TewWolk}.

    Recently, the ab-plane thermal conductivity of MgB$_2$ has been measured
as a function of magnetic field $H$ with orientations both parallel and
perpendicular to the c-axis \cite{Sologub2}. At low temperatures $\kappa(H)$
drops steeply for increasing $H$ up to a relatively low field which we denote
by $H^{(2)}_{c2}\,$ ($\sim$1 kOe), and then it rises continuously up to 
$H_{c2}\,$ ($\simeq$ 30 kOe for ${\bf H}\parallel{\bf c}$). The first drop of
$\kappa$ is interpreted as due to the behavior of $\kappa_{ph}(H)$ which is
caused by a strong suppression of the smaller gap $\Delta_2$ by relatively
small fields. This leads to a rapid increase in the number of those
quasiparticles that dominate the scattering of phonons \cite{Sologub1}. The
rapid suppression of the smaller gap can be explained in terms of a two-band
model with different energy gaps where the smaller gap is induced by
Cooper pair tunneling from the band with the larger gap \cite{Nakai}. 
The observed increase of $\kappa(H)$ above $H_{c2}^{(2)}$ is attributed to the 
increase of $\kappa_e(H)$ because $\kappa_{ph}$ has already reached its 
normal state value \cite{Sologub2}. Since this field dependence of 
$\kappa_e(H)$ at low temperatures is qualitatively similar to that of the 
specific heat coefficient $\gamma(H)\,$ \cite{Junod,Yang}, it is concluded 
that the growth of $\kappa_e(H)$ is due to the rapid increase of quasiparticles 
in the vortex cores associated with the larger energy gap. 

    In this Letter we present theories for the specific heat and the electronic
and phononic thermal conductivities in the vortex state for applied fields 
near the upper critical field $H_{c2}$. For simplicity we consider only one
isotropic s-wave pairing gap either on a cylindrical Fermi surface with 
${\bf H} \parallel {\bf c}\,$, or on a sperical Fermi surface. Our goal is to 
explain the measured quantities $\gamma(H)$ and $\kappa_e(H)$ near
$H_{c2}$ ($\simeq$ 30 kOe), and $\kappa_{ph}(H)$ below the effective 
upper critical field $H_{c2}^{(2)}$ ($\sim$ 1 kOe) for the vortex lattice
associated with the smaller energy gap. Our theories are based on the
normal and anomalous Green's functions $G$ and $F$ obtained from the
Gorkov integral equations with kernels given by the product of Abrikosov
vortex lattice order parameters $\Delta({\bf r_1})\Delta^{\ast}({\bf r_2})$ 
and the phase factor due to the magnetic field \cite{BPT}. These Green's 
functions depend sensitively on $\sin\theta$ where $\theta$ is the angle 
between the quasiparticle momentum ${\bf p}$ and the magnetic field 
${\bf H}$. For $\theta \rightarrow 0$, $G$ and $F$ tend to the Green's
functions for a BCS superconductor while, for $\theta \rightarrow \pi/2$, 
they take into account the effects of supercurrent flow and Andreev 
scattering due to the vortex lattice. This theory has been applied 
previously to calculate $\kappa_e(T,H)$ for energy gaps with line nodes 
like those occuring in the high-$T_c$ cuprates and Sr$_2$RuO$_4$ 
\cite{TewFay1}.

     A simplified version of the theory in Ref.~\onlinecite{BPT}, which has 
been derived from the Eilenberger equations \cite{Pesch}, yields the 
following expression for the spatial average of the density of states:
\begin{equation}
N(\omega,\theta)/N_0 \equiv A(\Omega;\tilde{\Delta},\theta) = 
\mbox{Re}\left[ 1+ \frac{8\tilde{\Delta}^2}{\sin^2\theta}
\left[ 1 + i\sqrt{\pi}\,z\,w(z) \right] \right]^{-1/2}\, .
\label{N}
\end{equation}
The quantities appearing in the equation are defined as follows:
$z = 2[\Omega\,\tilde{\Delta} + i(\Lambda/v)\,\Gamma]/\sin\theta\,$ with
$ \theta = \angle({\bf p},{\bf H})\,$, 
$\Lambda=(2eH)^{-1/2}\,$,
$ \tilde{\Delta}= \Delta\Lambda/v\,$,
$\Omega=\omega/\Delta\,$,
$\tilde{\Delta}^2=(H_{c2}-H)/6\beta_A H\,$,
$\Delta^2=\Delta^2_0(T)\left[ 1 - (H/H_{c2})\right]\,$,
and $w(z)=\exp(-z^2) {\mathrm{erfc}}(-iz) \,$.
Here $v$ is the Fermi velocity, $\beta_A$ the Abrikosov parameter which
we take as 1.2, $\Delta_0(T)$ is the BCS gap, and $\Gamma$ is the 
normal-state impurity scattering rate. In the Born and unitary scattering
limits, $\Gamma$ has to be replaced by $\Gamma\,A$ or $\Gamma/A\,$,
respectively, where $A(\Omega)$ is calculated self consistently. The 
specific heat C is given by
$C = N_0 T \int_0^{\infty}  dx \,x^2\,\mbox{sech}^2(x/2) A(Tx/\Delta) \,$.
At low temperatures ($T\ll \Delta$), $C=\gamma_s(H)T\,$, where
the specific heat coefficient 
$\gamma_s(H)$ is proportional to $A(\Omega=0)$. In Fig. 1 we have plotted 
$A(\Omega=0)$ versus $H/H_{c2}$ for $\sin\theta = 1$ corresponding to a 
cylindrical Fermi surface and ${\bf H}\parallel {\bf c}$, and impurity scattering
rates $\delta \equiv \Gamma/\Delta_0=\:$0.1, 0.2, and 0.5. A is seen to be 
nearly linear near $H_{c2}$ with a slope at $H_{c2}$ that decreases for 
increasing $\delta$. In Fig. 1 we also show our results for the angular 
average $\bar{A}=\int_0^{\pi/2}d\theta \sin\theta\,A(\theta)$ which corresponds
to a three dimensional Fermi surface. Comparison with the solid curves in
Fig. 1 shows that the slopes at $H_{c2}$ are decreased by the angular 
average. It should be pointed out that our results are only strictly valid in the
vicinity of $H_{c2}$. However, solution of the Eilenberger equations for a 
vortex lattice shows that the spatial average of the resulting density of states 
is well approximated by $A(\Omega)$ for fields down to about 
$(1/2)H_{c2}\,$ \cite{Dahm}.  The solid curve of $A$ vs $H/H_{c2}$ for 
$\delta=0.5$ in Fig. 1 agrees qualitatively with the measured field dependence
of the specific heat coefficient $\gamma_s(H)$ at low temperatures and fields 
near $H_{c2}\,$ \cite{Junod,Yang}. In the unitary impurity scattering limit $A$ 
increases slightly for decreasing field, however at lower fields our theory 
does not apply anyway.

     We turn now to the theory of the electronic thermal conductivity 
$\kappa_e$ in the vortex state near $H_{c2}$ which has been developed in 
Ref.~\onlinecite{TewFay1}. These expressions are easily modified to apply
to an isotropic s-wave pairing state and a field along the c-axis. Employing 
the expression valid at low temperatures we obtain for the ratio 
$\kappa_{es}/\kappa_{en}$ as a function of $H/H_{c2}$ the plots shown in
Fig. 2 for constant $\sin\theta = 1$ and impurity scattering rates 
$\delta=$0.1, 0.2, and 0.5. One sees that these plots exhibit upward
curvatures with decreasing slopes at $H_{c2}$ for increasing values of 
$\delta$. In Fig. 2 we have also plotted the corresponding results for the
angular averages over $\theta$ of $\kappa_{es}/\kappa_{en}$. Comparison
with the measured $\kappa(H)$ near $H_{c2}\,$, which is presumably 
dominated by the field dependence of $\kappa_e(H)\,$ \cite{Sologub2}, shows
that the measured upward curvature towards $H_{c2}$ is qualitatively best
described by the upper solid curve in Fig. 2 for the relatively large impurity
scattering rate $\delta=0.5$. It should be noted that the renormalization of the
impurity scattering rate in the unitary limit, $\Gamma \rightarrow \Gamma/A$,
slightly increases the upward curvature of $\kappa_{es}/\kappa_{en}$ in
comparison to that obtained without renormalization or in the Born scattering
limit. This is in somewhat better agreement with the data. Of course our 
theory cannot explain the measured field dependence of $\kappa_e(H)$ at
lower fields which first rises steeply and then saturates until the upward
curvature sets in at about $(1/2)\,H_{c2}$. Our theory also cannot describe 
the measured anomalous temperature dependence $\kappa_e \propto T^{1/2}$
for fields below approximately $(1/2)\,H_{c2}\,$: $\kappa_{en}$ and thus
$\kappa_{es}$ is proportional to $T/\Gamma$.

   It was argued in Ref.~\onlinecite{Sologub2} that the similar field
dependencies of the specific heat and electronic thermal conductivity can be
explained in terms of the relationship $\kappa_e = C_e v_F \ell/3$ where
$\ell$ is the mean free path. In fact, our theoretical expression for 
$\kappa_e$ has a similar form if one sets $\ell = v_F \tau_e$ where
$\tau_e = 1/\mbox{Im}\xi_0$ with $\mbox{Im}\xi_0 = \gamma + \gamma_A\,$
\cite{TewFay1}. Here $\gamma$ is the impurity scattering rate 
and $\gamma_A$ is the scattering rate due to Andreev scattering of the 
quasiparticles by the vortices. $\gamma_A$ is shown as a function of angle 
$\theta$ for different values of $\Omega$ and $\tilde{\Delta}$ in 
Ref.~\onlinecite{TewFay2}. These plots show that, for extended 
states ($\Omega \ge 1$),  $\gamma_A$ increases from zero 
as $\theta$ increases from 0 to $\pi/2$. This means that the Andreev 
scattering rate has a maximum for quasiparticles moving perpendicular to 
the vortex axis. For increasing field (decreasing $\tilde{\Delta}$), 
$\gamma_A$ decreases which has the effect that the curve for 
$\kappa_{es}/\kappa_{en}$ exhibits an upward curvature in comparison to 
$C_e$ which has a slight downward curvature for increasing field.

    Until now there exists only a phenomenological theory for the phonon 
heat transfer in the mixed state where the phonons are scattered by vortices
consisting of normal state cylinders \cite{Sousa}. We develop here the
microscopic theory of $\kappaê_{ph}$ (limited by electron scattering) in the
vortex state in close analogy to the theory for the BCS state \cite{BRT}. In 
that case the sum of probabilities for absorption and emission of phonons 
by quasiparticles yields a relaxtion rate $1/\tau_s$ for a phonon of energy
$\omega_0$ which is proportional to the integral over quasiparticle energy 
$E$ of the expression 
$|EE'/\varepsilon\varepsilon'|[1 - \Delta^2/EE'][f(E/T)-f(E'/T)]$. 
Here $E/\varepsilon=E/(E^2-\Delta^2)^{1/2}$ is the density of states, 
$E'=E + \omega_0$, and $f(E/T)$ is the Fermi function. In the mixed state 
near $H_{c2}$, the density of states $E/\varepsilon$ is replaced by 
$A(\Omega)$ as given in Eq. (\ref{N}). The coherence term 
$(\Delta/\varepsilon)(\Delta/\varepsilon')$, which arises from the matrix
elements for quasiparticle scattering, is replaced by $FF^{\dagger}$ where
the anomalous Green's function $F$ is given in Ref.~\onlinecite{TewFay1}. 
The spectral function of $F$ yields the following analog of 
$\Delta/\varepsilon\,$:
\begin{equation}
B(\Omega;\tilde{\Delta},\theta) = \mbox{Re} \left[
\frac{-i\,\sqrt{\pi}\,2\,(\tilde{\Delta}/\sin\theta)\,w(z)}
{ \left\{ 1 + (8\tilde{\Delta}^2/\sin^2\theta)[1+i\sqrt{\pi}\,z\, w(z)]
\right\}^{1/2} } \right]\, .
\label{B}
\end{equation}
The argument $z$ was defined below Eq.(\ref{N}). The
functions $A(\Omega)$ and $B(\Omega)$ are even and odd in $\Omega$.
For $\theta \rightarrow 0$ they tend to the BCS functions
$\omega/(\omega^2- \Delta^2)^{1/2}$ and $\Delta/(\omega^2- \Delta^2)^{1/2}$,
respectively. In this way we obtain for the ratio of phonon relaxation times 
in the normal and superconducting states (denoted by g in 
Refs.~\onlinecite{Sologub1}, \onlinecite{BRT}, and \onlinecite{TewWolk}):
\begin{eqnarray}
\tau_n/\tau_s = g(\Omega_0)
& = &
\left[ 1 - \exp(-\Omega_0\Delta/T)\right](2/\Omega_0)
\int_0^{\infty}d\Omega\,f[(\Omega-\Omega_0/2)\Delta/T]\,
f[-(\Omega+\Omega_0/2)\Delta/T] \nonumber\\
&  & \times
\left[ A(\Omega-\Omega_0/2) A(\Omega+\Omega_0/2)
- B(\Omega-\Omega_0/2) B(\Omega+\Omega_0/2) \right]\,.
\label{tauratio}
\end{eqnarray}
The quantity $\Omega_0 = \omega_0/\Delta$ is the phonon energy 
$\omega_0$ divided by the effective gap $\Delta=\Delta(H)\,$. 
For $T\ll\Delta$, the function $ff$
in the integrand of Eq.(\ref{tauratio}) is approximately one in the range 
from $\Omega=0$ to $\Omega_0/2$ and zero above $\Omega_0/2$. In 
Fig. 3 we show some examples of the function $g(\Omega_0)$ for parameter
values $\theta=\pi/2$, $\delta=0.1$, and $\tilde{\Delta}=$0.1, 0.2, 0.3, and 
0.6 corresponding to $H/H_{c2}=$0.93, 0.78, 0.61, and 0.28. For increasing
$\tilde{\Delta}$, or decreasing field, the function $g$ tends to the BCS step 
function which is zero in the range from $\Omega_0=0$ to 2, and $\pi/2$ for 
$\Omega_0>2\,$ (see Fig. 1 of Ref.~\onlinecite{TewWolk}). The physical 
meaning of $g$ at low temperatures for the BCS superconductor is that the
minimum energy of a phonon for creating a pair of quasiparticles is 
$\omega_0=2\Delta$. Fig. 3 shows that, for increasing field, the relaxation 
rate is more and more smeared out and the ratio $\tau_n/\tau_s$ tends to 
one. A similar effect occurs for increasing impurity scattering rate $\delta$. 

    The expression for the phonon thermal conductivity limited by electron
scattering and several other scattering processes is given by\cite{TewWolk}
\begin{equation}
\kappa_{ph} = A\,t^3\,\int_0^{\infty}dx\frac{x^4\,\mbox{e}^x}{(\mbox{e}^x-1)^2}
\left[ 1 + \alpha\,t^4  x^4 + \beta\,t^2 x^2 + \delta\,t x + \gamma\,t x\,g(xT/\Delta)
\right]^{-1}\, .
\label{Kappa}
\end{equation}
Here $x=\omega_0/T$, $t=T/T_c$, and the coefficients $A$, $\alpha$, 
$\beta$, $\delta$, and $\gamma$ refer to scattering by sample boundaries, 
point defects, sheet-like faults, dislocations, and quasiparticles, 
respectively. For $t\ll1$ and $T\ll\Delta$ the argument of $g(\Omega_0)$ in
Eq.(\ref{Kappa}) is approximately equal to zero. Taking the limit
$\Omega_0\rightarrow 0$ in Eq.(\ref{tauratio}) and noting that the coherence
term B vanishes at $\Omega_0=0$, we obtain 
$g(\Omega_0 = 0) = [A(\Omega_0=0, \tilde{\Delta}, \theta)]^2\,$.
For constant $\theta=\pi/2$ this expression yields the field dependence of
$\tau_n/\tau_s$ for the case ${\bf H}\parallel{\bf c}$ and a cylindrical Fermi 
surface. For a three dimensional Fermi surface, which is presumably more
appropriate for the quasiparticles associated with the smaller gap in
MgB$_2$, we have to multiply $[A(0,\theta)]^2$ by $\sin\theta$ and 
integrate over $\theta$ from 0 to $\pi/2$. In Fig. 4 we have plotted our
results for $g$ at $\theta=\pi/2$ and for the angular average of $g$ vs
$H/H_{c2}$ for impurity scattering rates $\delta$=0.1 and 0.5.

    The measured $\kappa_{ph}$ at $H=0\,$ \cite{Sologub1} has been 
analyzed in terms of an expression which is equivalent to our
Eq.(\ref{Kappa}) , apart from the term proportional to $\beta$. By 
inserting two different contributions for the electron scattering term 
$\gamma$ corresponding to different gaps $\Delta_1$ and $\Delta_2$, 
the fitting procedure indicated that the dominant part of the scattering of
phonons by electrons is provided by that part of the electronic excitation 
spectrum experiencing the smaller gap, $\Delta_2$. According to the 
two-band theory of Ref.~\onlinecite{Nakai}, this smaller gap is suppressed 
by a relatively small effective upper critical field $H_{c2}^{(2)}$ which leads 
to the measured fast drop of $\kappa_{ph}(H)$ as $H$ increases from 0 to
$H_{c2}^{(2)}\,$ \cite{Sologub2}. We conclude that the measured field
dependence of $\kappa_{ph}(H)$ below $H_{c2}^{(2)}$ at low
temperatures can be explained by inserting in Eq.(\ref{Kappa})
the values of the various constants obtained for $H=0$ and, for
$\tau_n/\tau_s = g(H)$, the angular average of 
$g(\Omega_0 = 0) = [A(\Omega_0=0, \tilde{\Delta}, \theta)]^2\,$. 
The behavior of $g(H)$, which determines the form of
the fast reduction of $\kappa_{ph}(H)$, depends on the impurity
scattering rate $\delta$ (see Fig. 4). We note that the model of
phonon scattering by vortices consisting of normal state  cores 
\cite{Sousa} yields a linear dependence of $g$ on $H/H_{c2}\,$.

     Our results have been derived for an isotropic s-wave gap. Recently
an anisotropic s-wave gap 
$\Delta({\bf p})/\Delta =(1+a\cos^2\theta)^{-1/2}$ has been proposed
which can explain the anisotropy of $H_{c2}$ in MgB$_2\,$
\cite{Posa}. For this gap with reasonable parameter values of
a = 10 and 20 we find that the effect of this gap anisotropy on the
field dependence of $C$ and $\kappa_{es}/\kappa_{en}$ is similar to 
that of impurity scattering. For instance, the curves of $C$ and 
$\kappa_{es}/\kappa_{en}$ versus $H/H_{c2}$ for $\delta = 0.1$
and $a=20$ appear to yield better fits to the experimental points
in Figs. 1 and 2 than the dashed curves for $a=0$ and $\delta=0.5\,$.

    The other important effect on the upward curvature of
$\kappa_{es}/\kappa_{en}$ arises from the decrease of the
Andreev scatering rate $\gamma_A$ with increasing field. 
This decrease is given approximately by the expression
$\gamma_A = \sqrt{\pi}(\tilde{\Delta}^2/\sin\theta)\mbox{exp}
[-(2\Omega\tilde{\Delta}/\sin\theta)^2]$ where 
$\tilde{\Delta}^2 \sim (H_{c2}-H)/H\,$. It is interesting that
the exponent in this expression, in particular the term
$(\sin\theta)^{-2}\,$, is similar to the exponent in the
scattering rate $1/\tau_v$ which has been derived previously
\cite{Yu}. In that model a quasiparticle is converted into a hole
due to Andreev reflection by vortex screening currents. This
process corresponds just to the imaginary part of the self
energy calculated in Ref.~\onlinecite{BPT}.

    Our theory is strictly valid only for fields near the upper 
critical field $H_{c2}\,$. However, recent numerical solutions
of the quasiclassical Eilenberger equations for the Abrikosov
vortex lattice \cite{Dahm} have shown that the analytical
expressions for the density of states \cite{BPT,Pesch}
(here the function $A(\Omega)$)  yield at low energies
impressively good results over the whole range of
fields down to $H_{c1}\,$. These results for low energies
are important for the calculation of thermodynamic
quantities. We expect that this is also true for the
function $B(\Omega)$, the spectral function of the
anomalous Green's function. 

    In conclusion, we have employed microscopic 
theories for the electronic specific heat $C(H)$
and the thermal conductivity $\kappa_e(H)$ and
developed a microscopic theory for the phononic
thermal conductivity $\kappa_{ph}(H)$ to explain
the measured field dependence of these quantities
in the mixed state of MgB$_2\,$. We find that impurity
and Andreev scattering, as well as Fermi surface
and gap anisotropy, give rise to large effects. For
sufficiently large impurity scattering rate, or gap
anisotropy, one can approximately fit the data
for $C(H)$ and $\kappa_e(H)$ for MgB$_2$ over
a broad field range below $H_{c2}\,$. Our new
theory for $\kappa_{ph}(H)$ is capable of
describing the observed rapid reduction of the
total $\kappa$ of MgB$_2$ for small fields if
this is actually caused by suppression of a
second smaller energy gap.

  We thank T. Dahm for helpful discussions.\\
\hrulefill\\
%

%
%
\newpage
\vspace{2cm}
\begin{figure}
\centerline{\psfig{file=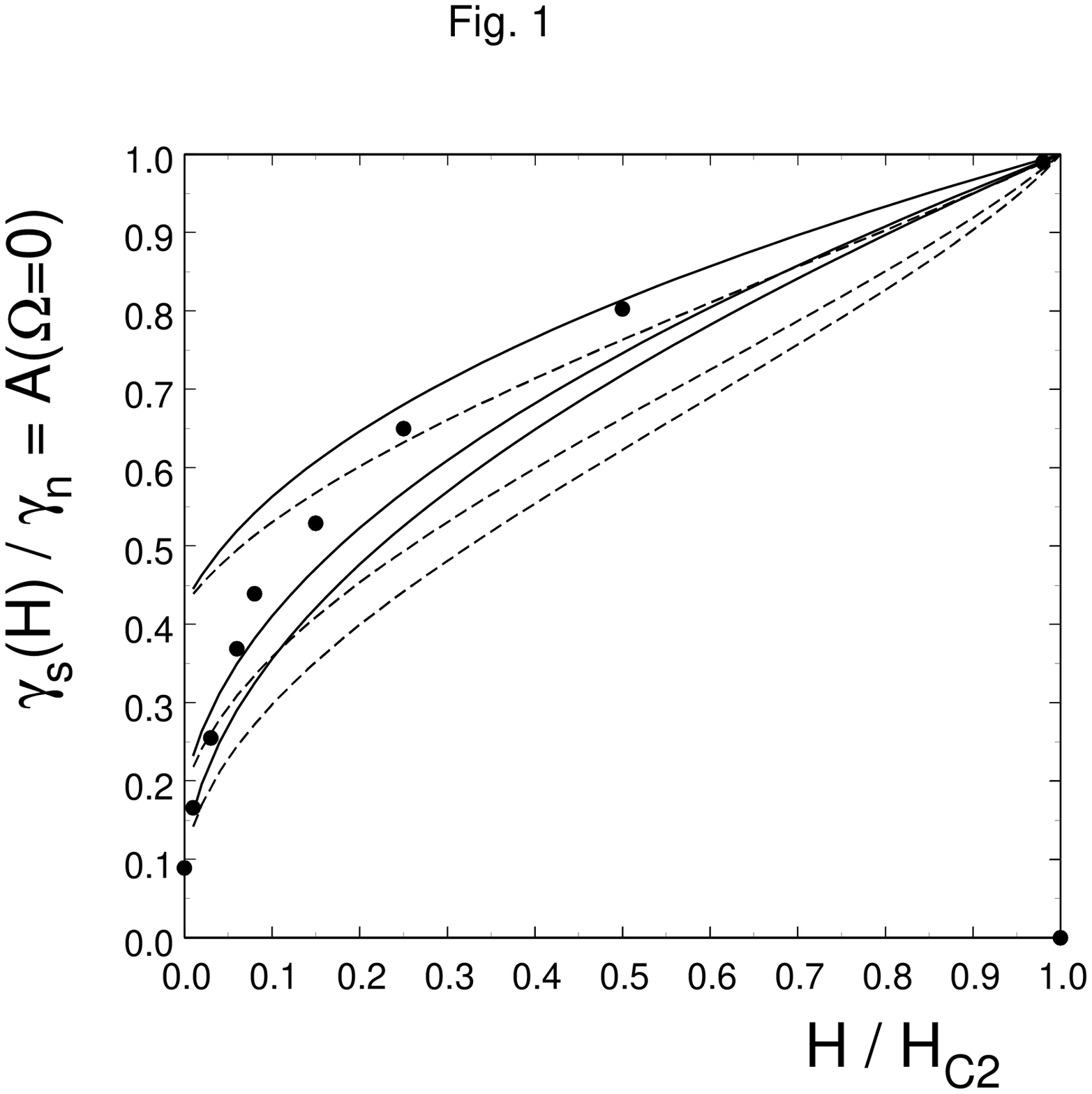,width=18cm,angle=0}}
\vskip -4cm
\caption{$A(\Omega=0,H) = \gamma_s(H)/\gamma_n$ vs $H/H_{c2}$ for
impurity scattering rates $\delta=\Gamma/\Delta_0=$0.1, 0.2, and
0.5  (from bottom to top). Solid curves for angle
$\theta=\pi/2$ corresponding to ${\bf H} \parallel {\bf c}$ for a
cylindrical Fermi surface ($\theta$ is the angle between the quasiparticle
momentum ${\bf p}$ and the field ${\bf H}$). Dashed curves for the average 
of $A$ over $\theta$ corresponding to a spherical Fermi surface. The
dots are reduced data from Ref. [8] for MgB$_2$ at $T=3K\,$. }
\label{fig1}
\end{figure}
\begin{figure}
\centerline{\psfig{file=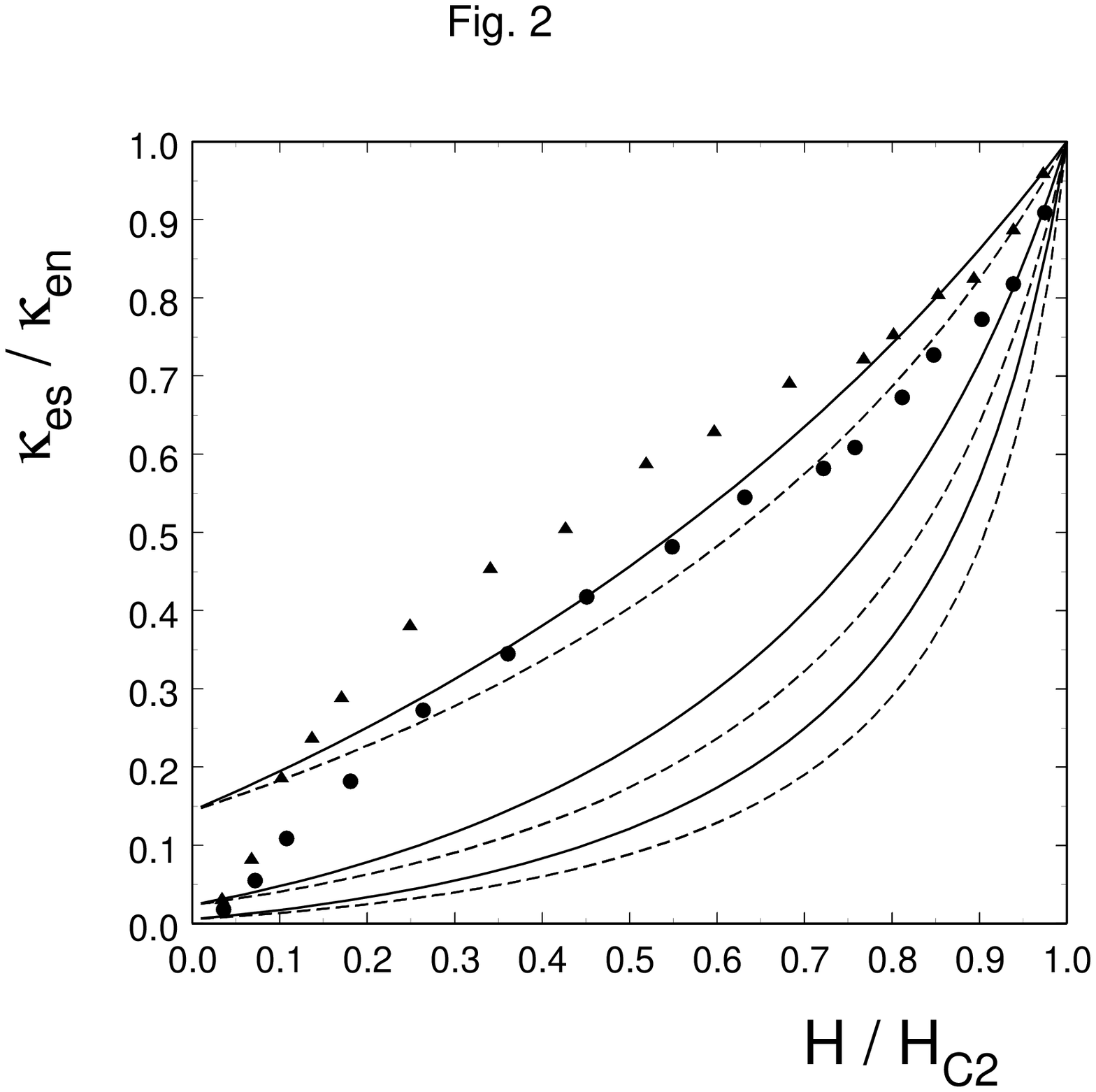,width=18cm,angle=0}}
\vskip -4cm
\caption{Electronic thermal conductivity ratio $\kappa_{es}/\kappa_{en}$ at low
temperatures vs $H/H_{c2}$ for impurity scattering rates $\delta$=0.1, 
0.2, and 0.5  (from bottom to top). Notation as in Fig.1: Solid curves for
angle $\theta=\pi/2$. Dashed curves for the average over $\theta$.
The dots (triangles) are reduced data from Ref. [6] for 
MgB$_2$ at $T=7.94K\,$ ($T=5.13K\, $) . }
\label{fig2}
\end{figure}
\begin{figure}
\centerline{\psfig{file=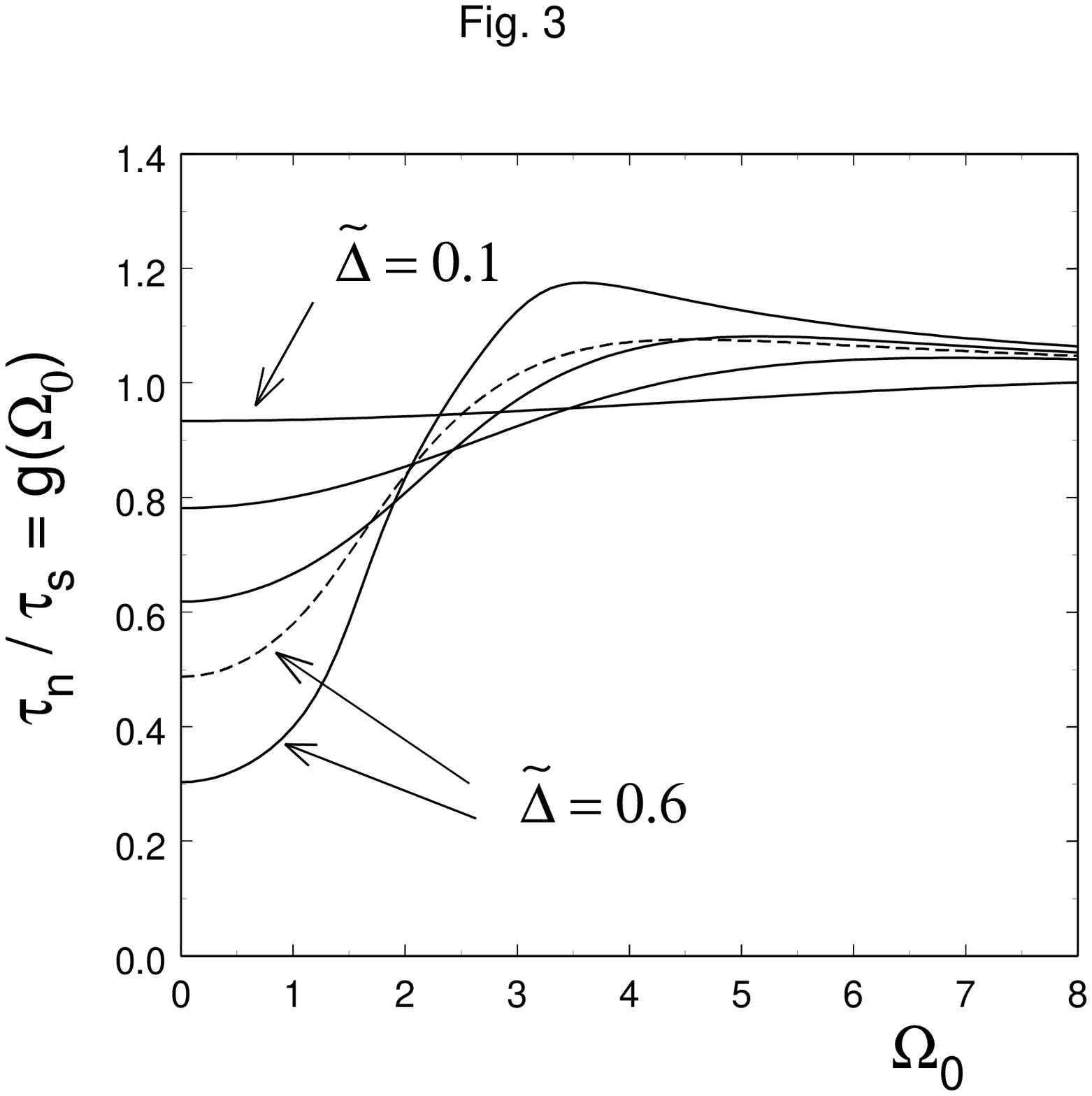,width=18cm,angle=0}}
\vskip -4cm
\caption{Ratio of phonon relaxation times $\tau_n/\tau_s = g(\Omega_0)$ due to 
electron scattering at low temperatures vs $\Omega_0 = \omega_0/\Delta$
where $\omega_0$ is the phonon energy and $\Delta=\Delta(H)$ the
effective energy gap. From top to bottom, $\tilde{\Delta}$=0.1, 0.2, 0.3, and
0.6, corresponding to $H/H_{c2}$= 0.93, 0.78, 0.61, and 0.28. Also, 
$\delta=0.1$ and $\theta=\pi/2$ (notation of Fig.1). The dashed curve
refers to $\tilde{\Delta}=0.6$ and impurity scattering rate $\delta=0.5\,$.}
\label{fig3}
\end{figure}
\begin{figure}
\centerline{\psfig{file=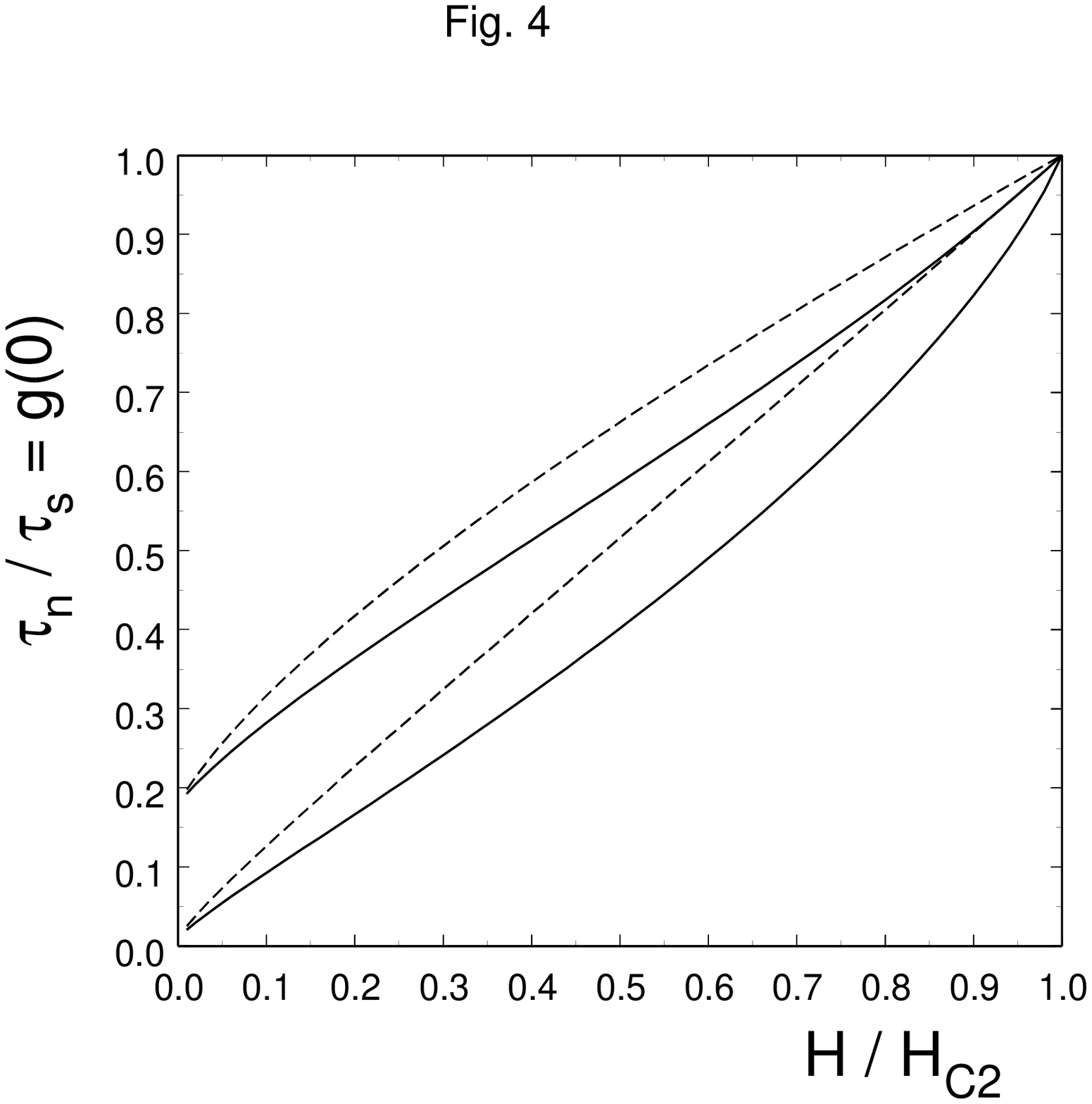,width=18cm,angle=0}}
\vskip -4cm
\caption{Ratio $\tau_n/\tau_s = g(\Omega_0=0,H)$ vs $H/H_{c2}$ for 
$\delta=0.1$ (lower curves) and $\delta=0.5$ (upper curves) with the 
notation of Fig.3. This quantity determines the reduction of the phonon 
thermal conductivity at low temperatures due to the increase of 
scattering by quasiparticles. The solid curves are the average 
of $g$ over $\theta$ corresponding to a spherical Fermi surface and
the dashed curves are $g$ for constant $\theta=\pi/2$. }
\label{fig4}
\end{figure}
\end{document}